\documentclass[sigconf]{acmart}

\usepackage{amsmath}
\usepackage{enumitem}
\usepackage{url}

\AtBeginDocument{%
  \setlength{\abovedisplayskip}{5pt plus 2pt minus 2pt}%
  \setlength{\belowdisplayskip}{5pt plus 2pt minus 2pt}%
  \setlength{\abovedisplayshortskip}{3pt plus 2pt}%
  \setlength{\belowdisplayshortskip}{4pt plus 2pt minus 2pt}%
}

\newcommand{\repoURL}{https://github.com/utshabkg/GCCP-reproduce}

\copyrightyear{2026}
\acmYear{2026}
\setcopyright{cc}
\setcctype{by}
\acmConference[CIKM '26]{Proceedings of the 35th ACM International Conference on Information and Knowledge Management}{November 07--11, 2026}{Rome, Italy}
\acmBooktitle{Proceedings of the 35th ACM International Conference on Information and Knowledge Management (CIKM '26), November 07--11, 2026, Rome, Italy}
\acmDOI{10.1145/3799682.3841055}
\acmISBN{979-8-4007-2539-5/2026/11}

\begin{document}

\title{When Do Anchor-Based Pointwise LLM Rerankers Help? Retriever Quality, Statistical Scope, and Anchor Design}

\author{Utshab Kumar Ghosh}
\affiliation{%
  \institution{Missouri University of Science and Technology}
  \city{Rolla} \state{MO} \country{USA}}
  \orcid{0000-0003-3096-6909}
  \email{u.ghosh@mst.edu}

\author{Shubham Chatterjee}
\affiliation{%
  \institution{Missouri University of Science and Technology}
  \city{Rolla} \state{MO} \country{USA}}
  \orcid{0000-0002-6729-1346}
  \email{shubham.chatterjee@mst.edu}

\renewcommand{\shortauthors}{Utshab Kumar Ghosh and Shubham Chatterjee}

\begin{abstract}
Anchor-based pointwise LLM reranking scores each candidate
against a shared reference passage to recover cross-document
context at pointwise cost. We study when this actually helps,
using GCCP/PAGC~\cite{long2025precise} as a representative
method. Our study is reproduction-first. We use reproduction as a starting point for a controlled component-level stress test of anchor-based pointwise reranking. Our initial reimplementation, based only
on the paper text, achieves $0.24$ nDCG@10 instead of the reported
$0.66$, revealing that several undocumented implementation details are
necessary to reproduce the method. After identifying and recovering eight
such details, we reproduce the reported results within $1.6\%$ and use the
validated implementation for controlled analysis.

We find that the core contrastive scoring idea is robust under rigorous
statistical correction. However, two design choices held fixed in the
original paper are less reliable. First, we find that combining the contrastive score with the standard
pointwise relevance score helps when the first-stage retriever is BM25,
but gives little or no benefit when the first-stage retriever is a
stronger dense model such as E5. Second, the paper's more complex method for constructing the
anchor is unnecessary. A much simpler anchor, built by interleaving the
top-ranked sentences, matches or outperforms it across datasets. These findings are consistent across different LLM backbones, including a
4-bit quantized 72B model. Overall, anchor-based pointwise reranking is
effective, but its gains come mainly from contrastive scoring rather than
from the more complex aggregation and anchor-construction choices, and
they appear under narrower conditions than the original evaluation
suggests.
\end{abstract}



\maketitle

\section{Introduction}
\label{sec:intro}

Zero-shot LLM rerankers are usually listwise, pairwise, or pointwise.
Pointwise methods are cheapest, with $\mathcal{O}(n)$ inferences for $n$
candidates, but they score each document independently. Long et~al.~\cite{long2025precise} propose GCCP/PAGC to address this.
Their method adds a query-focused \emph{anchor} passage to each
pointwise prompt, a short summary of the top-$m$ retrieved
candidates, so the LLM can score each candidate against a shared
reference. A standard yes/no relevance score is then averaged with
this contrastive score to produce the final ranking. In their paper,
this achieves listwise-competitive nDCG@10 at pointwise cost, without
the $\mathcal{O}(n^2)$ overhead of pairwise methods.

The original evaluation, however, holds three conditions fixed: BM25
as the first-stage retriever, a graph-partitioning algorithm (spectral
MDS) for anchor construction, and per-cell $t$-tests without
multiple-comparison correction. It remains unclear whether the reported gains persist with a stronger
first-stage retriever, a simpler anchor construction, or a more
conservative statistical protocol. These choices matter in practice:
a practitioner adopting the method needs to know not only whether it can
work, but also when it works and which components drive the gains.

We answer these questions through a \emph{reproduction-first}
methodology. We reimplement the full pipeline from the paper text
alone, validate within $1.6\%$ mean absolute nDCG@10 on TREC DL
2019/2020 and $1.9$--$4.5\%$ on eight BEIR datasets, and use that
validated platform for systematic boundary-condition analysis.

Our central finding is that anchor-based pointwise reranking is
\emph{effective but conditional}. The method has three components:
a standard yes/no relevance score (\textbf{RG-YN}), a contrastive
score that compares each candidate against the anchor
(\textbf{GCCP}), and an aggregation of both (\textbf{PAGC}).
Their contributions are not equal.

The full system (PAGC) reliably improves over standard pointwise
grading (RG-YN): Holm-significant in 12 of 22 settings, all
positive. The contrastive signal alone (GCCP vs RG-YN) is
directionally positive in 19 of 22 settings ($p \ll 0.001$,
sign test) but per-cell Holm-significant in only 3 of 22, useful but difficult to isolate at typical TREC DL query counts.

Aggregation adds value over GCCP alone in only 5 of 22 settings,
and is significantly \emph{harmful} on one dataset. This depends
strongly on the first-stage retriever. Under BM25, aggregation is consistently
beneficial. Under strong dense retrieval (E5), it is
statistically redundant on 7 of 8 BEIR datasets. The
graph-partitioning anchor construction is also unnecessary:
a simple composite of the top-$k$ candidate sentences matches
or outperforms it on every dataset we test. Our contribution is therefore not a new reranking architecture, but a controlled account of which parts of anchor-based pointwise reranking are load-bearing, which are incidental, and under what retrieval conditions the method remains useful.

\smallskip

\noindent Concretely, we address\footnote{Code and data at \url{\repoURL}.} the following research questions:

\begin{itemize}[leftmargin=2.5em,nosep]
    \item[\textbf{RQ1}] Can GCCP/PAGC be faithfully reproduced from
          the paper alone, and what does the process reveal about
          sensitivity to operational choices?

    \item[\textbf{RQ2}] Do the original statistical claims survive
          paired bootstrap testing with Holm-Bonferroni correction
          across multiple comparison families?

    \item[\textbf{RQ3}] How does first-stage retrieval quality
          moderate the value of anchor-based reranking and
          score aggregation?

    \item[\textbf{RQ4}] Is spectral MDS anchor construction necessary,
          or do simpler alternatives suffice?

    \item[\textbf{RQ5}] Does the mechanism transfer across LLM backbone
          families, including decoder-only and quantized models?
\end{itemize}

\smallskip

\noindent This paper makes three 
\textbf{contributions}:
\begin{itemize}[leftmargin=2.5em,nosep]
    \item A faithful reproduction of GCCP/PAGC from the paper 
          text alone, with a catalogue of eight undocumented 
          implementation choices including three that are 
          make-or-break.
    \item A controlled boundary-condition analysis showing when anchor-based pointwise reranking remains useful: its value depends strongly on first-stage retriever quality, the aggregation strategy, and anchor construction choices that the original evaluation held fixed.
    \item A methodological demonstration that per-cell uncorrected 
          $t$-tests overstate component contributions: 
          Holm-Bonferroni correction reduces apparent aggregation 
          significance from 8/22 to 5/22 settings, with one 
          significant negative result.
\end{itemize}

\section{Background and Related Work}
\label{sec:background}

We now describe GCCP/PAGC~\cite{long2025precise} in the form needed
for reproduction and controlled analysis, and situate this work in the reproducibility and statistical
evaluation literature. Given query $q$ and
candidate list $D{=}\{d_1,\ldots,d_n\}$, an anchor passage $d_a$
is constructed from the top-$m$ candidates via spectral
multi-document summarization: sentences are embedded with TF-IDF,
an affinity matrix is thresholded at $\theta$, the normalized graph
Laplacian is decomposed, and the Fiedler vector partitions sentences
into two clusters. The anchor is the larger cluster trimmed to $z$
sentences. The paper specifies $m{=}10$ and $z{=}10$. The threshold
$\theta$ is not stated. It uses three scoring components.

\emph{Relevance grading (RG-YN)} scores each candidate by yes/no
token probabilities:
\begin{equation}
f_{\text{RG-YN}}(q, d) =
    \frac{P(\texttt{yes} \mid q, d)}
         {P(\texttt{yes} \mid q, d) + P(\texttt{no} \mid q, d)}
\label{eq:rgyn}
\end{equation}
\emph{Contrastive scoring (GCCP)} scores each candidate against the
anchor via an A/B prompt:
\begin{equation}
f_c(q, d_i, d_a) =
    \frac{P(\texttt{A} \mid q, d_i, d_a)}
         {P(\texttt{A} \mid q, d_i, d_a) +
          P(\texttt{B} \mid q, d_i, d_a)}
\label{eq:gccp}
\end{equation}
\emph{Aggregation (PAGC)} averages both scores after per-query
min-max normalization:
\begin{equation}
f_{\text{PAGC}}(q, d) =
    \frac{1}{2}\!\left(\tilde{f}_{\text{RG-YN}}(q,d) +
    \tilde{f}_c(q, d, d_a)\right)
\label{eq:pagc}
\end{equation}
Several implementation details required to reproduce these equations
are not stated in the paper. We catalogue them in
Section~\ref{sec:sensitivity}.

Reproducibility studies in IR have repeatedly shown that unreported
implementation choices can dominate observed performance
differences~\cite{breuer2020how,lin2022buildingculturereproducibilityacademic,
yang2018anserini}, a concern amplified in LLM reranking pipelines
where retriever, prompt template, decoder configuration, and score
normalization each carry undocumented degrees of freedom. Without a
validated reimplementation, observed differences across conditions
could reflect implementation artifacts rather than method properties.

The paired $t$-test is typically applied per-cell in IR system
comparisons without correction for multiple comparisons, inflating
family-wise false positive rates~\cite{sakai2014statistical}. With
small query sets such as TREC DL ($n_q{=}43$--$54$), statistical
power is limited~\cite{sakai2006evaluating} and non-significant
results indicate absence of evidence rather than evidence of absence.
We use the Holm-Bonferroni procedure~\cite{holm1979} and paired
bootstrap testing~\cite{efron1994introduction,sakai2006evaluating}
throughout. Together, they provide a more conservative basis for our
boundary-condition analysis than per-cell uncorrected $t$-tests.
\section{Experimental Setup}
\label{sec:setup}

\subsection{Datasets}

We evaluate on TREC Deep Learning 2019~\cite{craswell2020overview}
and 2020~\cite{craswell2021overview} (43 and 54 queries respectively)
over MS MARCO v1 passages~\cite{bajaj2016msmarco}. For out-of-domain 
generalization we use the eight BEIR~\cite{thakur2021beir} subsets 
evaluated by Long et al.: TREC-COVID, Touché-2020, DBPedia-Entity,
SciFact, Signal1M, TREC-News, Robust04, and NFCorpus. Using the same 
subsets as the original paper allows direct comparison of our 
reproduction against their reported numbers.

\subsection{Retrieval}
We evaluate two primary first-stage retrievers. \textbf{BM25} is
run via Pyserini~\cite{lin2021pyserini}'s prebuilt Lucene indices
with parameters $k_1{=}0.9$, $b{=}0.4$. These values are not stated
in the original paper and are one of the operational choices we
identify in Section~\ref{sec:sensitivity}. \textbf{E5} uses
\texttt{intfloat/e5-base-v2}~\cite{wang2022e5} with top-100
retrieval via FAISS~\cite{johnson2019faiss} \texttt{IndexFlatIP}.
E5 is absent from the original paper. We include it to assess how
first-stage retrieval quality moderates reranker value (RQ3). We
additionally use \textbf{BGE} (\texttt{BAAI/bge-base-en-v1.5}~\cite{xiao2023cpack})
as a confirmatory dense retriever in Section~\ref{sec:retriever}
to verify that the DBPedia-Entity aggregation finding is not
specific to E5.

\subsection{Models}


We use the models used in the original GCCP paper: Flan-T5-Large (780M), Flan-T5-XL (3B), and Flan-UL2 
(20B), all in FP16. Additionally, we experiment with decoder-only models 
(Section~\ref{sec:decoder}): LLaMA-3.1-8B-Instruct, 
Qwen-2.5-7B-Instruct, 
Mistral-7B-Instruct-v0.3, and 
Qwen-2.5-72B-Instruct-AWQ (4-bit~\cite{lin2025awq}, single 48~GB 
GPU). Decoder-only models are prompted via
\texttt{tokenizer.apply\_chat\_template}. Details of the assistant
turn priming required for GCCP are discussed in
Section~\ref{sec:sensitivity}. All inference is sequential (no 
batching), matching the original code.

\subsection{Evaluation Metrics}


We report nDCG@10 throughout,
computed via \texttt{pytrec\_eval}~\cite{van2018pytreceval}, which
mirrors the reference \texttt{trec\_eval} binary. The nDCG implementation is not a minor detail. On TREC-COVID, a
hand-written nDCG routine differs from \texttt{pytrec\_eval} by about
2.5 points. Since several reported gains are of similar magnitude, this
choice can change whether a result appears meaningful.

\subsection{Statistical Testing}


We apply paired bootstrap 
significance testing~\cite{efron1994introduction,sakai2006evaluating} 
(10,000 resamples, seed 929) on per-query nDCG@10 deltas, with 95\% 
confidence intervals on the mean delta. To control the family-wise 
error rate across multiple comparisons we apply the Holm-Bonferroni 
step-down correction~\cite{holm1979} within each of three comparison 
families: PAGC vs RG-YN, PAGC vs GCCP-alone, and GCCP vs RG-YN. 
Each family covers $k{=}22$ primary settings: 14 TREC DL cells
(seven backbone$\times$retriever combinations across DL19 and DL20,
visualized in Figure~\ref{fig:holm}) and 8 BEIR cells
(Flan-T5-XL$\times$E5 across the eight BEIR subsets). We correct 
within families rather than jointly across all 66 tests because the 
three families test conceptually distinct hypotheses: whether 
contrastive anchor scoring helps at all (GCCP vs RG-YN), whether 
aggregation adds value beyond contrastive scoring alone (PAGC vs 
GCCP), and whether the full system improves over the pointwise 
baseline (PAGC vs RG-YN). Corrected p-values are denoted
$p_{\text{Holm}}$. Significance thresholds are $*$~($<$0.05),
$**$~($<$0.01), and $***$~($<$0.001). Per-query scores are released 
so readers can recompute under alternative correction procedures.

\subsection{Reproduction Scope}
\label{subsec:Reproduction Scope}


The original paper introduces
several PAGC variants combining GCCP with different pointwise
backbones (RG-YN, RG-S(0,4)~\cite{zhuang-etal-2024-beyond},
QG~\cite{sachan-etal-2022-improving}, and combinations thereof).
The strongest reported variant is PAGC\textsubscript{QSG} $=$
QG $+$ RG-S $+$ GCCP. We focus on the 2-component RG-YN $+$ GCCP
variant throughout, as it instantiates the paper's core
architectural idea with minimal confounds. We add a 3-component
PAGC-RS-YN-GCCP check in Section~\ref{sec:pagc_qsg} but do not
implement QG, which requires full-sequence query log-likelihood
and falls outside the scope of this study. We therefore use ``PAGC'' below to refer to the 2-component RG-YN+GCCP variant unless explicitly stated otherwise. Our claims about aggregation should be read as claims about this minimal, directly reproducible instantiation of the architecture, not as claims that every higher-order PAGC variant is unnecessary.
\section{RQ1: Reproduction and Sensitivity}
\label{sec:reproduction}

We address RQ1 by reimplementing the full GCCP/PAGC pipeline from the
paper text alone. We focus on the 2-component RG-YN$+$GCCP variant
defined in Section~\ref{subsec:Reproduction Scope}. After resolving the
implementation details described below, we match the reported numbers
closely: our mean absolute gap is 1.6\% on TREC DL and 1.9--4.5\% on
BEIR. This agreement gives us a faithful implementation for the
controlled analyses in Sections~\ref{sec:stats}--\ref{sec:transfer}.

During reproduction, we also identify substantial sensitivity to
operational choices that the original paper does not specify. We
catalogue eight such choices in Section~\ref{sec:sensitivity}. Together,
these choices explain the full nDCG@10 drop from $0.66$ to $0.24$ in our
initial paper-only reimplementation on DL19/Flan-T5-Large. The largest
single factor is the capitalization of the yes/no target tokens in the
relevance prompt. Changing this choice alone raises nDCG@10 from $0.24$
to $0.55$.

\subsection{Reproduction Results}
\label{sec:repro-results}

\subsubsection{Results on TREC DL 2019/2020}

Table~\ref{tab:dl-pagc} compares our reproduced PAGC results with the
reported TREC DL 2019 and 2020 results across three Flan-T5 model sizes.
Across the six settings, our nDCG@10 scores differ from the paper by only
1.6\% on average. The pipeline is deterministic within a single execution
(five-seed sweep, std $=0$), with sub-0.2 pt jitter across separate
executions that does not flip any conclusion in
Sections~\ref{sec:stats}--\ref{sec:transfer}.

The differences are small and do not point to a systematic implementation
error. We outperform the reported score in two settings
(DL19/Flan-T5-XL and DL20/Flan-T5-Large), remain within 1\% in two
others (DL20/Flan-T5-XL and DL20/Flan-UL2), trail by 1.5\% on
DL19/Flan-UL2, and fall below the paper by 2.5\% in the remaining
setting (DL19/Flan-T5-Large). We attribute this largest remaining gap to residual
preprocessing differences, which we discuss in
Section~\ref{sec:sensitivity}. Throughout this section, ``Paper'' refers
to the RG-YN+GCCP row of Table~1 in \cite{long2025precise}, which matches
our 2-component PAGC construction.

\begin{table}[t]
\centering
\caption{PAGC nDCG@10 on TREC DL. ``Paper'' values are
RG-YN+GCCP rows of Table~1 in \cite{long2025precise}; ``Gap''
is signed relative to ``Paper.'' Run-to-run jitter is
$\leq\pm 1$~pt nDCG@10 and does not flip any TREC-DL Holm
decision in Section~\ref{sec:stats}.}
\label{tab:dl-pagc}
\scalebox{0.85}{
\begin{tabular}{llccc}
\toprule
\textbf{Dataset} & \textbf{Model} & \textbf{Ours} & \textbf{Paper} & \textbf{Gap} \\
\midrule
DL19 & Flan-T5-Large & 0.6834 & 0.7012 & $-2.5\%$ \\
DL19 & Flan-T5-XL    & \textbf{0.7030} & 0.6969 & $+0.9\%$ \\
DL19 & Flan-UL2      & 0.7095 & 0.7206 & $-1.5\%$ \\
\midrule
DL20 & Flan-T5-Large & \textbf{0.6515} & 0.6281 & $+3.7\%$ \\
DL20 & Flan-T5-XL    & 0.6760 & 0.6810 & $-0.7\%$ \\
DL20 & Flan-UL2      & 0.7009 & 0.7023 & $-0.2\%$ \\
\bottomrule
\end{tabular}
}
\end{table}

\subsubsection{Results on BEIR}

Table~\ref{tab:beir-pagc} compares our reproduced PAGC nDCG@10 scores
with the reported BEIR results across eight datasets and three model
scales. The reproduction is closest for T5-XL, where our mean absolute
gap from the paper is 1.9\% and the median gap is 1.6\%. The largest gap
occurs for T5-Large, where the mean absolute gap is 4.5\%.

The gaps vary substantially across datasets and model scales. In several
cases, our reproduction outperforms the reported score, including
SciFact with T5-Large and Touché-2020 with T5-XL and UL2. In most other
cases, our scores are lower than the paper's. This mixed pattern does not
suggest a single systematic implementation error. Instead, it suggests
that the remaining differences likely come from dataset-specific
preprocessing choices.

The most likely sources are BEIR document-field handling, sentence
segmentation, and minor BM25 index differences. The original paper does
not specify, for each BEIR subset, whether documents use title, body, or
title plus body fields. Section~\ref{sec:sensitivity} also shows that
sentence segmentation can move PAGC scores, and small differences across
Pyserini prebuilt BM25 indices may further affect the candidate lists.
Identifying the exact source of each residual gap would require access to
the original preprocessing scripts, which are not included in the released
code. We therefore release our per-query scores to support future
diagnosis.

\begin{table}[t]
\centering
\caption{BEIR PAGC nDCG@10 across three model scales.}
\label{tab:beir-pagc}
\scalebox{0.8}{
\begin{tabular}{lcccccc}
\toprule
 & \multicolumn{2}{c}{\textbf{Flan-T5-Large}}
 & \multicolumn{2}{c}{\textbf{Flan-T5-XL}}
 & \multicolumn{2}{c}{\textbf{Flan-UL2}} \\
\cmidrule(lr){2-3}\cmidrule(lr){4-5}\cmidrule(lr){6-7}
\textbf{Dataset} & Ours & Paper & Ours & Paper & Ours & Paper \\
\midrule
SciFact           & \textbf{0.6403} & 0.6145 & 0.6840 & 0.6854 & 0.7114 & 0.7295 \\
NFCorpus          & 0.3620 & 0.3638 & 0.3728 & 0.3756 & 0.3776 & 0.3792 \\
TREC-COVID        & 0.7294 & 0.7641 & 0.7552 & 0.7761 & 0.7495 & 0.7781 \\
TREC-News         & 0.3820 & 0.4112 & 0.4490 & 0.4740 & 0.4563 & 0.4864 \\
Touch\'e-2020     & 0.2650 & 0.2928 & \textbf{0.3078} & 0.3016 & \textbf{0.3161} & 0.2930 \\
DBPedia           & 0.3898 & 0.4181 & 0.4061 & 0.4149 & 0.4138 & 0.4265 \\
Robust04          & 0.4800 & 0.4914 & 0.5279 & 0.5346 & 0.5347 & 0.5467 \\
Signal1M          & 0.2983 & 0.3027 & 0.3204 & 0.3230 & 0.3164 & 0.3186 \\
\midrule
\textbf{Avg.\ abs.\ gap} & \multicolumn{2}{c}{4.5\%}
                         & \multicolumn{2}{c}{1.9\%}
                         & \multicolumn{2}{c}{3.3\%} \\
\bottomrule
\end{tabular}
}
\end{table}

Together, the TREC DL and BEIR results show that our implementation is
faithful enough for controlled analysis. We reproduce the TREC DL results
within a 1.6\% mean absolute gap and obtain BEIR results that are close
overall, with dataset-specific residual differences that are consistent
with underspecified preprocessing rather than a major implementation
error. We therefore use this implementation as the basis for the boundary
condition analyses that follow. Importantly, the later analyses rely primarily on within-implementation contrasts: we compare RG-YN, GCCP, PAGC, retrievers, and anchor builders while holding preprocessing, candidate generation, prompts, scoring code, and evaluation fixed. Thus, residual paper-to-ours gaps affect exact comparability with Long et al., but they do not explain the main component-level conclusions reported in Sections~\ref{sec:stats}--\ref{sec:transfer}.

\subsection{Sensitivity to Operational Choices}
\label{sec:sensitivity}

Our paper-only reimplementation produced nDCG@10 $\approx 0.24$ on 
DL19 RG-YN against the original paper's 0.66. Consulting the authors' 
released code\footnote{\url{https://github.com/ChainsawM/GCCP}} 
revealed eight operational choices not stated in the paper that 
together account for the entire gap. We group them by severity.


\smallskip 
\noindent\textbf{Make-or-break choices.} Three choices individually produce a pipeline that runs to 
completion but yields broken rankings.

\begin{enumerate}[leftmargin=1.5em,nosep]

    \item \textbf{Decoder input string for T5.} The authors set 
          \texttt{decoder\_input\_ids} to \texttt{`<pad>~'} for 
          RG-YN and \texttt{`<pad>~Passage~'} for GCCP. With an 
          empty decoder input, T5 produces near-uniform token 
          probabilities, rendering both scores uninformative.

    \item \textbf{Target-token case for RG-YN.} The correct tokens 
          are lowercase \texttt{`yes'/`no'}, not 
          \texttt{`Yes'/`No'}. With a prompt ending in 
          ``Answer `yes' or `no'\,'' essentially all probability 
          mass concentrates on the lowercase variant. This single 
          choice moves nDCG@10 on DL19 RG-YN from 0.24 to 0.55
          in isolation.

    \item \textbf{Per-query min-max normalization before 
          aggregation.} Equation~11 of \cite{long2025precise} 
          writes the PAGC score as a plain average. The authors' 
          released code (\path{src/utils/aggregate_utils.py}) 
          inserts a per-query, per-component min-max normalization 
          to $[0,1]$ before averaging. Without it, RG-YN's $[0,1]$ 
          probabilities and GCCP's contrastive logits live on 
          incomparable scales, and aggregation reduces to 
          ``whichever component has larger raw magnitude.''

\end{enumerate}

\smallskip

\noindent\textbf{Performance-relevant choices.}
Five further choices do not individually collapse the pipeline but 
materially affect results.

\begin{enumerate}[leftmargin=1.5em,nosep,resume]

    \item \textbf{Target-token case for GCCP.} Uppercase 
          \texttt{`A'/`B'}, with the prompt ending ``Output 
          Passage A or Passage B:''. Case flips relative to 
          RG-YN (item~2) and is not stated in the paper.

    \item \textbf{Spectral threshold $\theta{=}0.2$.} Not stated. 
          With $\theta{=}0$ the affinity graph is dense and the 
          Fiedler partition becomes uninformative.

    \item \textbf{BM25 parameters $k_1{=}0.9$, $b{=}0.4$.} Not 
          stated. Pyserini's stock defaults ($k_1{=}0.82$, 
          $b{=}0.68$) change the candidate pool composition.

    \item \textbf{Document truncation to 128 tokens.} Not stated. 
          Without it, long candidates overflow the encoder and 
          the prompt template no longer fits.

    \item \textbf{nDCG implementation.} The paper does not specify
          which nDCG routine to use. A hand-written implementation
          produces a $\sim$2.5~pt difference on TREC-COVID relative
          to \texttt{pytrec\_eval}, large enough to flip
          conclusions in cells where reported gains are of similar
          magnitude.

\end{enumerate}

\subsubsection{Implications}

The two-group structure matters for anyone attempting to reimplement 
this or similar pipelines: the three make-or-break choices must be 
recovered exactly, while the five performance-relevant choices 
require care but allow some variation. More broadly, each of these
choices \emph{fails silently}. The model produces plausible-looking
probabilities, the pipeline runs to completion, and the output is
wrong. This is not unique to GCCP. It reflects a known gap between
published method descriptions and the implementations that produce
reported results in neural ranking
pipelines~\cite{breuer2020how,lin2022buildingculturereproducibilityacademic,yang2018anserini}. 
We release our full catalogue and per-query scores for independent 
verification.
\section{RQ2: Do the Statistical Claims Survive Correction?}
\label{sec:stats}

The original paper reports paired $t$-tests for
``$+$~GCCP vs single component'' only, without multiple-comparison
correction or PAGC-vs-GCCP-alone tests. This leaves it unclear whether
PAGC improves because of the contrastive score, the aggregation step, or
both. We therefore rerun the analysis with Holm-Bonferroni correction
and test each component separately.

We extend this analysis in three ways. First, we use paired bootstrap tests on per-query nDCG@10 differences. We use two-sided $p$-values. We choose paired bootstrap testing because it makes fewer assumptions about the distribution of per-query scores than a paired $t$-test, while retaining comparable power in IR evaluation
\cite{sakai2006evaluating}. Second, we test all 22 primary settings
defined in Section~\ref{sec:setup}. Third, we test three comparison
families: GCCP vs RG-YN, PAGC vs GCCP, and PAGC vs RG-YN.

Each comparison answers a different question. GCCP vs RG-YN tests whether
contrastive anchor scoring helps on its own. PAGC vs GCCP tests whether
combining the contrastive score with the pointwise score adds value beyond
contrastive scoring alone. PAGC vs RG-YN tests whether the full method
improves over standard pointwise grading.

\begin{figure}[t]
\centering
\includegraphics[width=0.93\columnwidth]{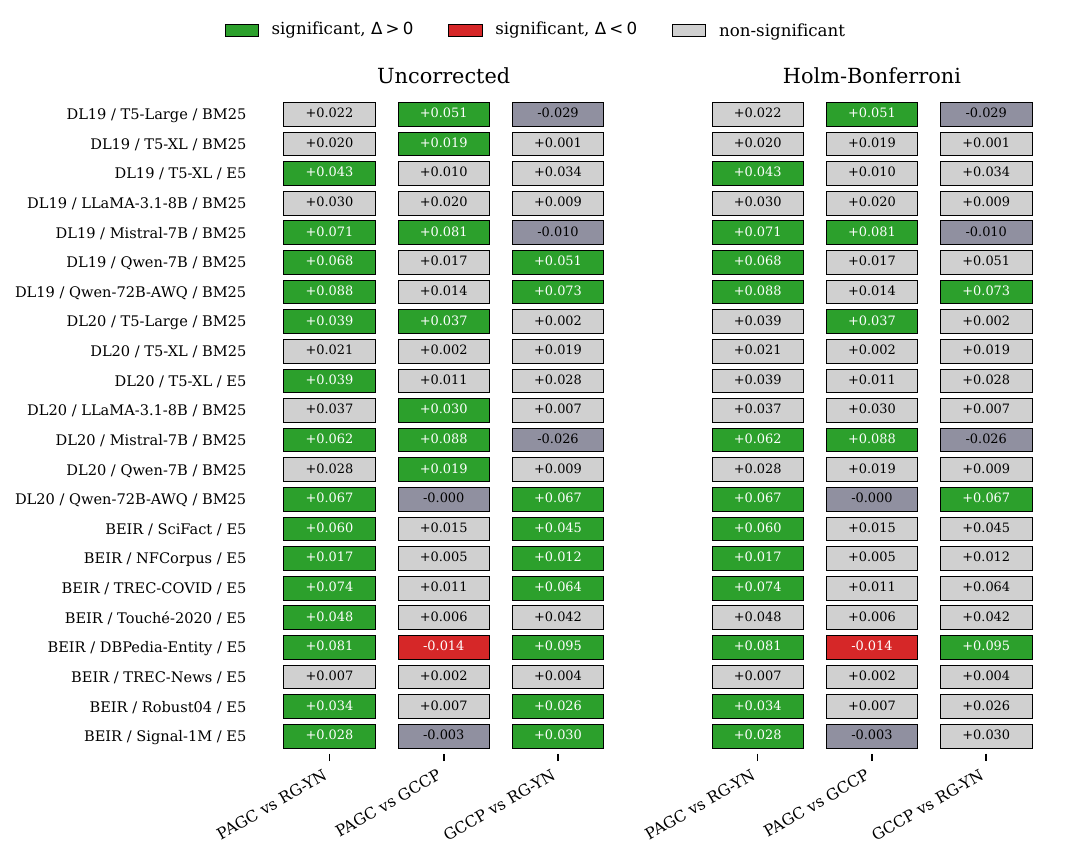}
\caption{Significance of pairwise comparisons across all 22 primary 
settings, before and after Holm-Bonferroni correction. Each cell 
shows the mean nDCG@10 delta; colour indicates significance status 
(green: $p_{\text{Holm}}{<}0.05$, $\Delta{>}0$; red: 
$p_{\text{Holm}}{<}0.05$, $\Delta{<}0$; grey: non-significant). 
Correction reduces PAGC-vs-GCCP significance from 8 to 5 settings; 
the single red cell (DBPedia-Entity, PAGC vs GCCP, $\Delta{=}{-}0.014$) 
is the only significant negative result.}
\Description{Two side-by-side heat-map panels, titled ``Uncorrected'' and
``Holm-Bonferroni''. Each panel has 22 rows, one per primary setting: 14 TREC DL
backbone-by-retriever cells across DL19 and DL20, then the eight BEIR datasets
under E5 retrieval. Each panel has three columns: PAGC vs RG-YN, PAGC vs GCCP,
and GCCP vs RG-YN. Every cell prints a signed mean nDCG@10 delta and is shaded
green when significant and positive, red when significant and negative, and grey
when non-significant. The counts of significant cells fall from 15, 8 and 9 in
the uncorrected panel to 12, 5 and 3 in the corrected panel. A single red cell,
DBPedia-Entity under PAGC vs GCCP, appears in both panels.}
\label{fig:holm}
\end{figure}

For each comparison family, we run 22 tests: 14 TREC DL settings
(seven backbone--retriever combinations evaluated on both DL19 and DL20,
as enumerated in Figure~\ref{fig:holm}) and 8 BEIR settings
(Flan-T5-XL with E5 retrieval across the eight BEIR subsets). Because the
three families test different hypotheses, we apply Holm-Bonferroni
correction~\cite{holm1979} separately within each family of 22 tests,
rather than jointly across all 66 tests. To make this choice auditable, our artifact includes the raw bootstrap $p$-values and adjusted $p$-values under both within-family and global Holm correction. The qualitative conclusion is unchanged: PAGC improves over RG-YN more reliably than it improves over GCCP alone.

The correction changes the picture. For PAGC vs RG-YN, the number of
significant results drops from 15 of 22 to 12 of 22. For PAGC vs GCCP, it
drops from 8 of 22 to 5 of 22. For GCCP vs RG-YN, it drops from 9 of 22
to 3 of 22. Figure~\ref{fig:holm} shows the full pattern.

The strongest result is for the full method. PAGC remains significantly
better than RG-YN after Holm correction in 12 of 22 settings. All 12
significant differences are positive. The significant gains appear on six
of the eight BEIR datasets with E5 first-stage retrieval, and on 6 of the
14 TREC DL settings, mostly with larger LLM backbones. Thus, adding
anchor-contrastive scoring to standard pointwise grading provides a real
benefit, and this benefit survives correction.

The aggregation step is less reliable. PAGC significantly improves over
GCCP alone in only 5 of 22 settings. Four of these significant gains are
positive, and all four occur with BM25. One significant result is
negative: on DBPedia-Entity under E5, PAGC is worse than GCCP alone
($\Delta{=}{-}0.0144$, $p_{\text{Holm}}{=}0.032$). The E5 results are
especially important. Under E5, aggregation is either non-significant or
negative in every setting. This suggests that aggregation depends on the
quality of the first-stage retriever. We examine this pattern in
Section~\ref{sec:retriever}.

The contrastive score alone shows a weaker per-setting result. GCCP alone
is significantly better than RG-YN after Holm correction in only 3 of 22
settings. These are two Qwen-72B-AWQ settings and DBPedia-Entity under
E5. In the other 19 settings, GCCP still improves the mean score, but the
improvement is too small to survive correction within each individual
setting. This is especially likely on small TREC-DL datasets, which have
only $43$--$54$ queries. A sign test over the 22 mean differences confirms
that the overall trend is positive ($p \ll 0.001$).

These results should be interpreted with sample size in mind. TREC DL 2019 has
only $n_q{=}43$ queries whereas DL 2020 has $n_q{=}54$ queries. With so few queries, a non-significant
Holm-corrected result does not necessarily mean that there is no effect.
It may also mean that the test does not have enough power to detect a
small effect \cite{sakai2014statistical}. We therefore treat non-significant results on small datasets as absence
of evidence, not evidence of absence. The BEIR datasets have more queries
and provide more statistical power. In these E5-based BEIR settings, we
can more clearly detect the negative DBPedia-Entity result and the lack of
consistent benefit from aggregation.
\section{RQ3: How Does First-Stage Retrieval Quality Moderate
Reranker Value?}
\label{sec:retriever}

We test a boundary condition for anchor-based reranking: does
it still help when the first-stage retriever is already strong? The
original paper evaluates GCCP/PAGC only after BM25 retrieval. This matters
because anchor-based reranking may be most useful when the first-stage
candidate list is noisy. If the retriever already places many relevant
documents near the top, the reranker has less remaining error to correct,
and the aggregation step may add less value.

We test this by replacing BM25 with E5-base-v2~\cite{wang2022e5} as the
first-stage retriever. We retrieve the top 100 documents using
FAISS~\cite{johnson2019faiss} \texttt{IndexFlatIP}. We keep the reranking
pipeline unchanged. This isolates the role of first-stage retrieval
quality: the reranker stays the same, but the candidate list changes.

The main result is that stronger first-stage retrieval sharply reduces
the marginal value of reranking. With BM25 on DL20, PAGC improves over the
retriever alone by $+0.197$ nDCG@10. With E5, the same reranker improves
over the retriever alone by only $+0.013$. On SciFact, PAGC with E5 even
falls below the E5 baseline ($-0.010$). Thus, the key result is not simply
that E5 is stronger than BM25. The important point is that PAGC has much
less residual work to do once the first-stage candidate list is already
strong.

\begin{figure}[t]
\centering
\includegraphics[width=0.93\columnwidth]{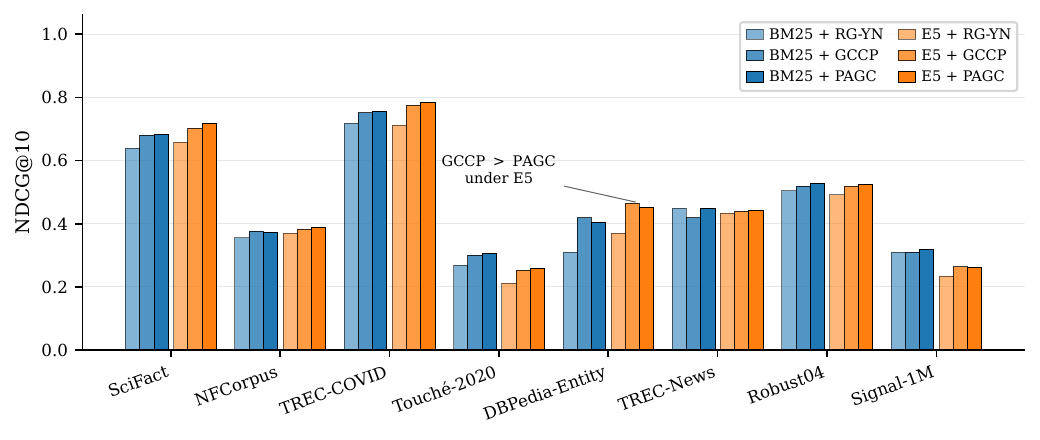}
\caption{nDCG@10 on eight BEIR datasets with Flan-T5-XL reranking. Under BM25
(blue), anchor-based scoring usually helps and PAGC is often strongest.
Under E5 (orange), gains shrink and aggregation is less reliable; on
DBPedia-Entity, GCCP alone beats PAGC.}
\Description{Grouped bar chart of NDCG@10, on a vertical axis from 0.0 to 1.0,
for the eight BEIR datasets SciFact, NFCorpus, TREC-COVID, Touch\'e-2020,
DBPedia-Entity, TREC-News, Robust04 and Signal-1M. Each dataset has six bars:
three blue bars for BM25 first-stage retrieval with RG-YN, GCCP and PAGC, and
three orange bars for E5 first-stage retrieval with RG-YN, GCCP and PAGC. An
annotation reading ``GCCP $>$ PAGC under E5'' points to the DBPedia-Entity
group.}
\label{fig:retriever}
\end{figure}

The first-stage scores support this interpretation. On TREC DL, E5
improves first-stage nDCG@10 over BM25 by $+20$ points on DL19 and
$+23$ points on DL20. This stronger retrieval baseline changes the
comparison with more complex rerankers. At the T5-XL scale, PAGC built on E5 retrieval
outperforms both the paper's 2-component RG-YN$+$GCCP setup on BM25
($+2.2$ points on DL19 and $+3.7$ points on DL20) and the paper's
3-component PAGC-QSG setup on BM25
($+1.2$ points on DL19 and $+3.0$ points on DL20). In these settings,
improving the first-stage retriever gives larger gains than adding more
reranking components after BM25. Figure~\ref{fig:retriever} shows this pattern across all eight BEIR
datasets at the Flan-T5-XL scale.

\begin{table}[t]
\centering
\caption{8-set BEIR generalization under E5 retrieval + Flan-T5-XL 
rerank. Bold marks the per-dataset winner. Significance markers are 
from paired bootstrap testing (10{,}000 resamples, seed 929) with 
Holm-Bonferroni correction ($k=22$ within each comparison family): 
\textbf{ns}: $p_{\text{Holm}}\geq 0.05$; *: $<0.05$; **: $<0.01$; 
***: $<0.001$.}
\label{tab:beir-e5}
\scalebox{0.7}{
\setlength{\tabcolsep}{4pt}
\begin{tabular}{lcccccc}
\toprule
\textbf{Dataset} & \textbf{$n_q$} & \textbf{RG-YN} & \textbf{GCCP} & \textbf{PAGC}
& \textbf{$\Delta$ PAGC$-$GCCP} & \textbf{$\Delta$ PAGC$-$RG-YN} \\
\midrule
SciFact          & 300 & 0.6579 & 0.7026 & \textbf{0.7176} & $+0.0150$~ns   & $+0.0597$~*** \\
NFCorpus         & 323 & 0.3712 & 0.3836 & \textbf{0.3884} & $+0.0048$~ns   & $+0.0172$~*** \\
TREC-COVID       &  50 & 0.7117 & 0.7752 & \textbf{0.7859} & $+0.0107$~ns   & $+0.0742$~**  \\
Touch\'e-2020    &  49 & 0.2118 & 0.2538 & \textbf{0.2594} & $+0.0056$~ns   & $+0.0477$~ns  \\
DBPedia-Entity   & 400 & 0.3707 & \textbf{0.4659} & 0.4515 & $-0.0144$~\textbf{*} & $+0.0808$~*** \\
TREC-News        &  57 & 0.4346 & 0.4391 & \textbf{0.4415} & $+0.0024$~ns   & $+0.0069$~ns  \\
Robust04         & 249 & 0.4920 & 0.5184 & \textbf{0.5255} & $+0.0071$~ns   & $+0.0335$~*** \\
Signal1M        &  97 & 0.2334 & \textbf{0.2638} & 0.2612 & $-0.0025$~ns   & $+0.0279$~**  \\
\midrule
\textbf{Avg.\ (8 sets)} & -- & 0.4354 & 0.4753 & \textbf{0.4789} & $+0.0036$ & $+0.0435$ \\
\bottomrule
\end{tabular}
}
\end{table}

Table~\ref{tab:beir-e5} gives three component-level findings. First, the
full PAGC method still improves over standard pointwise grading with E5
retrieval. PAGC is better than RG-YN in all 8 BEIR datasets, and the
difference is Holm-significant in 6 of 8. The two non-significant cases
are Touché-2020 ($n_q{=}49$) and TREC-News ($\Delta{=}{+}0.007$). Both
differences are still positive, so the non-significance is consistent
with limited power or small effect size rather than a reversal.

Second, aggregation does not add reliable value under E5. PAGC does not
Holm-significantly improve over GCCP alone on any of the 8 datasets. On
DBPedia-Entity, the largest dataset ($n_q{=}400$), PAGC is significantly
worse than GCCP alone
($\Delta_{\text{PAGC-GCCP}}{=}{-}0.0144$,
$p_{\text{Holm}}{=}0.032$). This negative result does not appear to be
only a large-$n_q$ effect. Robust04 is the next-largest E5 setting
($n_q{=}249$), but it shows a small positive and non-significant
difference
($\Delta_{\text{PAGC-GCCP}}{=}{+}0.0071$). This suggests that the DBPedia-Entity result is not explained by statistical power alone. One plausible factor is the entity-centric nature of the task, but identifying the exact dataset property responsible requires further analysis.

Third, GCCP alone shows a positive but weaker pattern against RG-YN.
GCCP is better than RG-YN in all 8 datasets, but the difference is
Holm-significant in only one dataset: DBPedia-Entity
($p_{\text{Holm}}{<}0.001$). This matches the broader pattern from
TREC DL. The contrastive anchor usually helps, but the effect is often too
small to isolate within a single dataset after correction.

We next ask why aggregation stops helping under E5. One possible
explanation is that RG-YN and GCCP become more similar when the
first-stage retriever is stronger. To test this, we compute per-query
Kendall's~$\tau$ between RG-YN and GCCP scores across the 100 candidates,
and then average within each setting. Mean~$\tau$ is $0.43$ in BM25
settings and $0.47$ in E5 settings. This difference is in the expected
direction, but it is modest. It helps explain why aggregation gains shrink
under E5, but it does not explain the DBPedia-Entity negative result.
DBPedia-Entity has a mid-range agreement value ($\tau{=}0.48$), so score
agreement alone does not identify it as unusual.

The DBPedia-Entity result also holds with a different dense retriever. We
replace E5 with BGE-base-en-v1.5~\cite{xiao2023cpack} and rerun the
Flan-T5-XL reranking pipeline on three BEIR datasets. We treat this as a
separate confirmatory analysis and apply Holm-Bonferroni correction within
this family of three tests. This does not change the primary $k{=}22$
correction used in Section~\ref{sec:stats}.

Under BGE, PAGC is again significantly worse than GCCP alone on
DBPedia-Entity. The effect is larger than under E5
($\Delta_{\text{PAGC-GCCP}}{=}{-}0.0190$,
$p_{\text{Holm}}{<}0.001$, compared with $-0.0144$ under E5). The other
two datasets do not show this negative pattern. Robust04 has a small
positive and non-significant difference under BGE
($\Delta_{\text{PAGC-GCCP}}{=}{+}0.0014$), similar to E5
($+0.0071$, non-significant). NFCorpus is also positive and
non-significant under BGE ($+0.0055$), again similar to E5
($+0.0048$). As a positive control, PAGC remains significantly better
than RG-YN on all three BGE datasets.

The BGE and E5 first-stage scores are similar on these datasets:
$0.4040$ vs $0.4234$ on DBPedia-Entity, $0.4383$ vs $0.4391$ on Robust04,
and $0.3722$ vs $0.3517$ on NFCorpus. This does not prove that the two
retrievers return the same candidate lists, but it does show that they
provide first-stage rankings of similar effectiveness. The repeated
DBPedia-Entity pattern therefore does not appear to be specific to E5.

The practical implication is that PAGC is not uniformly worth running.
With BM25, PAGC is useful: aggregation is Holm-significant in 4 of 12 BM25
settings, and the gain over the first-stage retriever is large. With the strong dense first-stage retrievers we test, the picture changes. Under E5, and in our limited BGE confirmation, GCCP alone is usually competitive with PAGC. Aggregation adds no reliable
benefit and can hurt on entity-heavy datasets such as DBPedia-Entity.
\section{RQ4: Is Spectral MDS Anchor Construction Necessary?}
\label{sec:anchor}

This section tests whether the most engineered part of GCCP is actually
responsible for its gains. GCCP builds its anchor using spectral
multi-document summarization. This procedure embeds sentences with TF-IDF,
constructs an affinity matrix, decomposes the normalized graph Laplacian,
and partitions sentences using the Fiedler vector. This is substantially
more complex than simply using ranked passages or sentences as the anchor.

We therefore compare spectral MDS against three simpler anchor builders:
a random candidate passage, the top-1 BM25 passage, and a top-3
sentence-interleaved composite. We find that this added complexity is not
justified. Spectral MDS is never the best anchor in our TREC DL or BEIR
ablations. Simpler anchors consistently match or outperform it. This suggests that GCCP
does not depend on spectral anchor construction. It mainly needs a
reasonable contrastive reference text.

\smallskip

\noindent\textbf{TREC DL anchor ablation.}
%
%
Table~\ref{tab:anchor-full} reports GCCP and PAGC nDCG@10 on DL19 and
DL20 with Flan-T5-Large. We vary only the anchor builder.

\begin{table}[t]
\centering
\caption{Anchor-builder ablation on full DL19/DL20 with
Flan-T5-Large. RG-YN is identical across rows by construction
(0.6634 / 0.6121) and omitted. \textbf{Spectral MDS is not the
strongest anchor in our reproduction.}}
\label{tab:anchor-full}
\small
\begin{tabular}{lcc|cc}
\toprule
& \multicolumn{2}{c}{\textbf{DL19}} & \multicolumn{2}{c}{\textbf{DL20}} \\
\textbf{Anchor builder} & GCCP & PAGC & GCCP & PAGC \\
\midrule
Random passage   & 0.6394 & 0.6945 & 0.6103 & 0.6474 \\
Top-1 BM25       & \textbf{0.6511} & \textbf{0.6948} & 0.6131 & 0.6485 \\
Top-3 composite  & 0.6410 & 0.6947 & \textbf{0.6280} & \textbf{0.6572} \\
Spectral MDS (paper) & 0.6341 & 0.6852 & 0.6137 & 0.6507 \\
\bottomrule
\end{tabular}
\end{table}

On TREC DL, spectral MDS is not the strongest anchor. On DL19, the Top-1
BM25 passage beats spectral MDS by $+1.7$ points for GCCP and $+1.0$
point for PAGC. On DL20, the Top-3 sentence-interleaved composite is best,
beating spectral MDS by $+1.4$ points for GCCP and $+0.7$ points for
PAGC. Averaged across DL19 and DL20, GCCP with Top-1 reaches $0.6321$,
compared with $0.6239$ for GCCP with Spectral.

This result agrees with the direction reported in the original paper.
Table~5 of Long et al.~\cite{long2025precise} also shows Top slightly outperforming
Spectral on TREC DL: GCCP$+$Top reaches $0.6099$, while GCCP$+$Spectral
reaches $0.6076$. Our reproduction confirms this exception and finds a
larger difference. The implication is not that spectral MDS fails
catastrophically. Rather, a simple high-ranking passage already provides
a strong enough anchor in these experiments, leaving little room for the
more complex spectral method to show an advantage.


\smallskip

\noindent\textbf{BEIR anchor ablation.}
Table~\ref{tab:beir-anchor} reports GCCP nDCG@10 across all 8 BEIR
datasets with Flan-T5-Large and BM25 retrieval. We vary only the anchor
builder. PAGC follows the same pattern and is omitted for space.

\begin{table*}[t]
\centering
\caption{GCCP nDCG@10 across all 8 BEIR datasets with
Flan-T5-Large / BM25, varying only the anchor builder. Bold marks
the per-dataset winner. Top-3 composite wins 5/8, Random wins
3/8, Spectral wins 0/8. ``Ours avg.''\ is the simple mean
across 8 datasets; ``Paper avg.''\ is the corresponding number
from \cite{long2025precise} Table~5 (paper does not include a
``Top-3 composite'' baseline).}
\label{tab:beir-anchor}
\scalebox{0.90}{
\setlength{\tabcolsep}{4pt}
\begin{tabular}{lcccccccc|cc}
\toprule
\textbf{Anchor} & SciFact & NFCorpus & TREC-COVID & Touch\'e & Robust04 & TREC-News & Signal1M & DBPedia & \textbf{Ours avg.} & \textbf{Paper avg.} \\
\midrule
Random           & 0.5781 & 0.3213 & \textbf{0.7701} & \textbf{0.2954} & 0.4386 & \textbf{0.4384} & 0.2894 & 0.4075 & 0.4423 & 0.4253 \\
Top-1 BM25       & 0.6659 & 0.3303 & 0.7609 & 0.2894 & 0.4423 & 0.4218 & 0.2734 & 0.4255 & 0.4512 & 0.4346 \\
Top-3 composite  & \textbf{0.6692} & \textbf{0.3381} & 0.7649 & 0.2869 & \textbf{0.4505} & 0.4261 & \textbf{0.3020} & \textbf{0.4268} & \textbf{0.4581} & --     \\
Spectral (paper) & 0.6195 & 0.3311 & 0.7564 & 0.2694 & 0.4380 & 0.4245 & 0.2990 & 0.4186 & 0.4446 & \textbf{0.4471} \\
\bottomrule
\end{tabular}
}
\end{table*}

The BEIR results are clearer than the TREC DL results. Spectral MDS is
not the best anchor on any of the 8 datasets. It finishes last on three
datasets: TREC-COVID, Touché-2020, and Robust04. On the remaining five
datasets, it finishes second or third. The aggregate ordering in our
results is Top-3 $>$ Top-1 $>$ Spectral $>$ Random. This differs from the
aggregate ordering reported in the original paper, where Spectral is the
strongest anchor.

These results show that the spectral construction is not load-bearing. A
simple anchor built from top-ranked sentences is at least as effective,
and often better, across the datasets we test.

\smallskip 

\noindent\textbf{Checking the ``Top'' baseline.}
The original paper reports a ``Top'' anchor baseline, but does not fully
specify how this anchor is constructed. For example, it is unclear whether
``Top'' means the full top-ranked BM25 passage, only its title, or some
other text field. This matters because a mismatch in this baseline could
make our simpler-anchor comparison misleading.

To rule out this possibility, we test eight variants of the ``Top'' and
``Random'' anchors (Table~\ref{tab:beir-anchor-variants}). This lets us
identify which variant most closely matches the paper's reported baseline
and check whether our conclusion about Spectral MDS depends on this
choice.

\begin{table*}[t]
\centering
\caption{Anchor-operationalization sweep on all 8 BEIR sets,
Flan-T5-Large/BM25 (4 representative datasets shown for the
per-cell columns; full 8-cell results in the released artifact).
\textbf{Top-1 title-only matches the paper's reported ``Top''
average ($0.4313$ vs $0.4346$, $\Delta{-}0.003$)}, identifying the
likely operationalization the paper uses. Random-5 averaging
partially explains the Random gap. Spectral matches the paper
within $0.003$ regardless of the other variants tested.}
\label{tab:beir-anchor-variants}
\small
\setlength{\tabcolsep}{4pt}
\begin{tabular}{lrrrrr|l}
\toprule
\textbf{Variant} & \textbf{Avg.} & SciFact & DBPedia & Touch\'e & Signal1M
& \textbf{vs paper Table~5} \\
\midrule
Random-1 (our default)        & 0.4423 & 0.5781 & 0.4075 & 0.2954 & 0.2894 & paper Random $0.4253$, $\Delta{+}0.017$ \\
Random-5 averaging            & 0.4364 & 0.5673 & 0.4115 & 0.2805 & 0.2825 & $\Delta{+}0.011$ (closes $\sim 35\%$) \\
\midrule
Top-1 BM25 (full doc, default) & 0.4512 & 0.6659 & 0.4255 & 0.2894 & 0.2734 & paper Top $0.4346$, $\Delta{+}0.017$ \\
Top-3 sentence composite      & \textbf{0.4581} & 0.6692 & 0.4268 & 0.2869 & 0.3020 & not in paper \\
Top-5 sentence composite      & 0.4547 & 0.6476 & 0.4259 & 0.2888 & 0.3029 & not in paper \\
Top-1 64-char truncation      & 0.4157 & 0.5798 & 0.3815 & 0.2653 & 0.2671 & undershoots by $\Delta{-}0.019$ \\
\textbf{Top-1 title-only}     & \textbf{0.4313} & 0.6499 & 0.3776 & 0.2728 & 0.2750 & \textbf{$\Delta{-}0.003$ (matches)} \\
\midrule
Spectral MDS (paper builder)  & 0.4446 & 0.6195 & 0.4186 & 0.2694 & 0.2990 & paper $0.4471$, $\Delta{-}0.003$ \\
\bottomrule
\end{tabular}
\end{table*}

This check does not change the conclusion. Top-1 title-only matches the
paper's reported ``Top'' average within $0.003$ ($0.4313$ vs $0.4346$),
which makes it the most likely operationalization used in the paper.
Spectral also matches the paper closely ($0.4446$ vs $0.4471$). Thus, the
negative result is not caused by a poor reproduction of the spectral
anchor or by an obviously mismatched Top baseline.

After this check, spectral MDS still does not finish first on any of the
8 BEIR datasets. It is also beaten on aggregate by both the Top-3
sentence-interleaved composite ($0.4581$) and the full-document Top-1
anchor ($0.4512$). The extra engineering in spectral MDS is therefore not
justified by the results.

For Random, our variants reduce but do not fully eliminate the gap to the
paper. Random-5 averaging closes about $35\%$ of the difference. The
remaining gap likely reflects a different random-sentence choice that we
could not identify. We contacted the corresponding author but did not
receive a response before submission. This uncertainty does not affect the
main conclusion, because Spectral itself matches the paper closely and
still loses to simpler anchors in our ablations.


\smallskip

\noindent\textbf{Hyperparameter sensitivity.}
We also test whether the spectral anchor is sensitive to its
hyperparameters. On DL20/T5-Large/BM25, we sweep
$m \in \{5, 10, 15, 20\}$, $z \in \{5, 10, 15, 20\}$, and
$\theta \in \{0.1, 0.2, 0.3, 0.4\}$. PAGC nDCG@10 varies by about
$\pm1.5$ points across $m$ and $\theta$. This is a meaningful range,
because many reported gains on these benchmarks are around two points.

This sensitivity strengthens the practical concern. Spectral MDS is not
only more complex than simpler anchors. It also introduces tunable
hyperparameters that can materially affect performance. Practitioners
should tune $m$ and $\theta$ on a held-out split rather than copying the
paper's defaults. In contrast, $z$ is less important beyond $z{=}10$,
because the encoder's 512-token limit truncates additional anchor
sentences. Thus, the paper's choice of $z{=}10$ is reasonable, but it is
not the source of the method's gains.

Overall, the anchor ablation points to a simple conclusion: GCCP benefits
from having a contrastive anchor, but the spectral MDS procedure is not
necessary. A much simpler anchor built from top-ranked passages or
sentences gives comparable or better performance with less engineering
and fewer tunable choices.
\section{RQ5: Does the Mechanism Transfer Across Backbone 
Families?}
\label{sec:transfer}

The previous sections established \emph{when} anchor-based
reranking helps: under noisy first-stage retrieval, with
sufficient query count, and with simple anchor construction.
This section asks whether those patterns hold when the backbone
changes entirely. The original paper evaluates only
encoder-decoder Flan models. We extend to four decoder-only
instruction-tuned LLMs, including a 4-bit AWQ-quantized 72B
model running on a single 48~GB GPU.

\subsection{Decoder-Only LLMs}
\label{sec:decoder}

We adapt the RG-YN and GCCP prompts to chat-template format 
via \path{tokenizer.apply_chat_template(..., 
add_generation_prompt=True)}, reading next-token probabilities 
at the assistant turn boundary. For GCCP, the assistant turn
is primed with the literal string \texttt{`Passage '} before
reading logits, mirroring the T5
\texttt{decoder\_input\_text=`<pad>~Passage~'} convention and
avoiding a $4\times$ collapse in nDCG@10 we verified on a
Qwen-2.5-0.5B smoke test. We run four models: 
LLaMA-3.1-8B-Instruct~\cite{llama3report}, 
Qwen-2.5-7B-Instruct~\cite{qwen25technical}, 
Mistral-7B-Instruct-v0.3~\cite{jiang2023mistral}, and 
Qwen-2.5-72B-Instruct-AWQ (4-bit, single 48~GB GPU).
Table~\ref{tab:decoder} reports nDCG@10 on TREC-DL. Three
findings stand out.

\begin{table}[t]
\centering
\caption{nDCG@10 on TREC DL with decoder-only LLMs adapted via
chat templates (novel; not in the original paper). Bold marks
the strongest GCCP and PAGC cells. Comparison Flan numbers
(Table~\ref{tab:dl-pagc}): T5-Large $0.6834/0.6515$;
T5-XL $0.7030/0.6760$; UL2 $0.7095/0.7009$; paper UL2 RG-YN+GCCP
$0.7206/0.7023$ (Table~1 of \cite{long2025precise}).}
\label{tab:decoder}
\small
\scalebox{0.90}{
\begin{tabular}{llcccc}
\toprule
\textbf{Dataset} & \textbf{Model} & \textbf{BM25} & \textbf{RG-YN} & \textbf{GCCP} & \textbf{PAGC} \\
\midrule
DL19 & LLaMA-3.1-8B-Instruct    & 0.5058 & 0.6427 & 0.6521 & 0.6723 \\
DL19 & Qwen-2.5-7B-Instruct     & 0.5058 & 0.6527 & 0.7039 & 0.7212 \\
DL19 & Mistral-7B-Instr.\ v0.3  & 0.5058 & 0.5716 & 0.5617 & 0.6429 \\
DL19 & Qwen-2.5-72B-AWQ         & 0.5058 & 0.6590 & \textbf{0.7323} & \textbf{0.7465} \\
\midrule
DL20 & LLaMA-3.1-8B-Instruct    & 0.4796 & 0.5795 & 0.5866 & 0.6166 \\
DL20 & Qwen-2.5-7B-Instruct     & 0.4796 & 0.6360 & 0.6447 & 0.6641 \\
DL20 & Mistral-7B-Instr.\ v0.3  & 0.4796 & 0.5360 & 0.5105 & 0.5983 \\
DL20 & Qwen-2.5-72B-AWQ         & 0.4796 & 0.6311 & \textbf{0.6983} & \textbf{0.6980} \\
\bottomrule
\end{tabular}
}
\end{table}

\textbf{The contrastive anchor transfers to large quantized 
models.}
Qwen-2.5-72B-AWQ is the strongest model we tested. On DL19
PAGC it reaches $0.7465$, surpassing both our reproduced
Flan-UL2 ($0.7095$) and the paper's Flan-UL2 ($0.7206$) by
more than 2.5~pts. A 4-bit open-weight 72B model on a
single 48~GB GPU is competitive with the paper's best
dense-FP16 result. We frame this as a scaling data point
spanning a $\sim$3.6$\times$ parameter-count gap (72B vs
20B), not a PAGC-method gain. On DL20 it matches our reproduced Flan-UL2 within $0.003$
($0.6980$ vs $0.7009$), with PAGC and GCCP essentially tied here
($0.6980$ vs $0.6983$), echoing Section~\ref{sec:retriever}: aggregation
adds less value once the underlying scorer is strong. This confirms
that 4-bit AWQ quantization is not a barrier to the mechanism.

\textbf{The DBPedia-Entity negative survives backbone 
change.}
On a compute-capped subset of the first 50 queries on
DBPedia-Entity ($\sim$4~min/query at 72B, with first-$N$ by qid
order carrying the usual $n_q{=}50$ sampling variance), the
negative finding from Section~\ref{sec:retriever} reproduces:
RG-YN($0.3667$) $<$ PAGC($0.4139$) $\approx$ E5($0.4168$)
$<$ GCCP($\mathbf{0.4380}$). A paired bootstrap (10,000
resamples) confirms PAGC$<$GCCP ($\Delta{=}{-}0.024$,
$p_{\text{raw}}{=}0.043$). PAGC and the E5 first-stage are
statistically tied ($\Delta{=}{-}0.003$,
$p_{\text{raw}}{=}0.93$). The DBPedia-Entity negative
therefore survives a complete change of backbone family
(encoder-decoder $\to$ decoder-only), parameter count
(3B $\to$ 72B), and quantization regime (FP16 $\to$ 4-bit
AWQ). This strengthens the claim from
Section~\ref{sec:retriever} that the result reflects a
property of the retrieval setting (entity-retrieval
semantics under strong dense retrieval) rather than an
artefact of the Flan-T5 backbone.

\textbf{At 7--8B, backbone family matters more than 
parameter count.}
Qwen-2.5-7B is competitive with Flan-T5-XL on DL19 PAGC
($0.7212$ vs $0.7030$), and the difference is not statistically
significant at $n{=}43$. LLaMA-3.1-8B 
falls below Flan-T5-Large ($-1.1$/$-3.5$~pts on DL19/DL20) 
despite being $10\times$ larger by parameter count. 
Mistral-7B-v0.3 underperforms on GCCP specifically (A/B 
discrimination $0.5617$/$0.5105$), suggesting it does not 
respond reliably to the contrastive prompt format. The 
Qwen-2.5 vs Mistral gap at DL19 PAGC ($+0.078$) exceeds 
the Flan-T5-Large vs Flan-UL2 gap ($+0.026$), confirming 
that backbone family is at least as important as parameter 
count in this regime. On SciFact the ordering from 
Section~\ref{sec:retriever} reproduces cleanly at 72B: 
E5($0.7993$) $<$ RG-YN($0.8074$) $<$ GCCP($0.8586$) $<$ 
PAGC($\mathbf{0.8734}$).

\subsection{Aggregation Strategy}
\label{sec:agg}


We test whether equal-weight linear aggregation is the best
way to combine RG-YN and GCCP. We compare the default
$\alpha=0.5$ average with an $\alpha$-weighted sweep and with
Borda, Condorcet, and Copeland rank voting across eight
dataset--retriever--model cells. Alternative aggregation rules do
not materially change the result: Borda wins in 3/8 cells, but its
mean score is essentially tied with the default linear average
(0.6544 vs.\ 0.6556).

The main exception appears under E5 retrieval. In all four E5
cells, the best weighted variant is $\alpha=0.25$, which gives more
weight to GCCP and less to RG-YN. This is consistent with RQ3:
when the first-stage retriever is stronger, RG-YN appears to add
less useful residual signal, so equal weighting can dilute the
contrastive score. The gains are small, so equal-weight aggregation
remains a reasonable default.

The rank-voting results are also limited. With only two input
rankers, Borda, Condorcet, and Copeland are equivalent in our
setting, so they cannot distinguish positional aggregation
strategies. The narrower conclusion is that changing the
aggregation rule does not rescue aggregation under E5. The larger
factor is how much independent signal RG-YN contributes beyond
GCCP.

\subsection{Three-Component Aggregation}
\label{sec:pagc_qsg}

We add the RG-S$(0,4)$ component to form a 3-component
PAGC-RS-YN-GCCP variant (Table~\ref{tab:pagc_rs}). QG is
left for future work as it requires full-sequence query
log-likelihood.
\begin{table}[t]
\centering
\caption{PAGC-RS-YN-GCCP (paper's Table~4 ``RG-S+RG-YN+GCCP''
homogeneous variant) on Flan-T5-Large/BM25. Our TREC average
$(0.6856 + 0.6325)/2 = 0.6591$ vs the paper's $0.6634$, a
0.4-point gap, consistent with our other reproductions.}
\label{tab:pagc_rs}
\scalebox{0.8}{
\begin{tabular}{lcc|cc}
\toprule
& \multicolumn{2}{c}{\textbf{Ours (T5-Large/BM25)}}
& \multicolumn{2}{c}{\textbf{Paper (Table 4 hom.)}} \\
\textbf{Method}        & DL19 & DL20 & TREC avg & BEIR avg \\
\midrule
RG-S only              & 0.5872 & 0.5204 & --     & --     \\
RG-YN only             & 0.6634 & 0.6121 & --     & --     \\
GCCP only              & 0.6341 & 0.6137 & --     & --     \\
PAGC = RG-YN $+$ GCCP  & 0.6852 & 0.6507 & --     & --     \\
PAGC = RG-S $+$ GCCP   & 0.6547 & 0.6058 & --     & --     \\
PAGC-RS-YN-GCCP        & 0.6856 & 0.6325 & 0.6634 & 0.4538 \\
\bottomrule
\end{tabular}
}
\end{table}

The 3-component variant reproduces the paper's Table~4 number
within 0.4 points on TREC average. On DL19 it is essentially 
tied with 2-component PAGC ($0.6856$ vs $0.6852$). On DL20 
it \emph{loses} ($0.6325$ vs $0.6507$, $-1.8$~pts): RG-S 
alone is substantially weaker than RG-YN on DL20 ($0.5204$ 
vs $0.6121$), and aggregating a weak component drags the 
mean rank. The value of adding RG-S is therefore non-monotonic
across datasets, a nuance hidden by the paper's choice to
report only TREC$+$BEIR averaged. This echoes the RQ3 finding 
that aggregation benefit depends on component 
complementarity: when a component's errors overlap with the 
existing components, aggregation hurts rather than helps.

\subsection{Positioning Against Listwise and Pairwise Baselines}
\label{sec:rankgpt}

Table~\ref{tab:rankgpt} compares our reproduced 2-component PAGC with
the listwise and pairwise baselines reported in the original paper's
Table~2. We do not re-run these baselines, so we use them only as
paper-reported reference points.

\begin{table}[t]
\centering
\caption{Our reproduced 2-component PAGC compared with the
listwise and pairwise baselines reported in Table~2 of
\cite{long2025precise}. Baseline numbers are cited from the
paper and are not independently re-run.}
\label{tab:rankgpt}
\small
\scalebox{0.90}{
\begin{tabular}{lcc}
\toprule
\textbf{Method / Backbone} & \textbf{DL19} & \textbf{DL20} \\
\midrule
BM25 (no rerank) & 0.5058 & 0.4796 \\
\midrule
\multicolumn{3}{l}{\emph{Paper-reported baselines from Table~2 of \cite{long2025precise}}} \\
\quad RankGPT~\cite{sun-etal-2023-chatgpt} / Flan-T5-Large          & 0.6690 & 0.6260 \\
\quad PRP-Allpair~\cite{qin-etal-2024-large} / Flan-T5-Large      & 0.6646 & 0.6201 \\
\quad PRP-Heapsort~\cite{qin-etal-2024-large} / Flan-T5-Large     & 0.6570 & 0.6190 \\
\quad PRP-Graph-40~\cite{qin-etal-2024-large} / Flan-T5-Large     & 0.6690 & 0.6270 \\
\quad Setwise-Heapsort~\cite{zhuang2024set} / Flan-T5-Large & 0.6700 & 0.6180 \\
\quad PAGC-QSG / Flan-T5-Large (3-comp) & \textbf{0.6973} & 0.6433 \\
\midrule
\multicolumn{3}{l}{\emph{Our reproduced 2-component PAGC}} \\
\quad PAGC / Flan-T5-Large             & 0.6852 & \textbf{0.6507} \\
\bottomrule
\end{tabular}
}
\end{table}

Our 2-component PAGC outperforms all paper-reported listwise and pairwise
baselines on both DL19 and DL20. It remains below the paper's
3-component PAGC-QSG on DL19, but exceeds it on DL20. This places the
simpler 2-component version in the same performance range as the stronger
reranking baselines reported by the original paper.

\subsection{Inference Cost}

We measure wall-clock latency on DL19 with 100 candidates per query on an
NVIDIA RTX~6000 Ada. PAGC costs roughly the sum of its two scoring passes:
Flan-T5-Large takes 2.08 seconds/query for RG-YN, 2.39 for GCCP, and 4.47
for PAGC. Flan-T5-XL takes 1.99, 2.65, and 4.64 seconds/query,
respectively. This is expected because RG-YN and GCCP do not share
LLM-side computation, although the two passes could in principle be
parallelized. Flan-UL2 is substantially slower, taking about 23.3
seconds/query end-to-end. Anchor construction is not the bottleneck:
spectral MDS contributes less than 1\% of total pipeline time. Thus, the
main reason to replace spectral MDS with a top-3 sentence-interleaved
composite is simplicity and effectiveness, not latency.

\section{Discussion}
\label{sec:discussion}

\noindent\textbf{When do anchor-based pointwise rerankers help?}
Our results support a more conditional view of anchor-based pointwise
reranking. The core idea is useful: adding an anchor-based contrastive
score to standard pointwise grading produces robust gains after
statistical correction. However, the gains do not come equally from every
part of the original pipeline. The contrastive score is the most reliable
component. The aggregation step and the spectral MDS anchor are less
stable.

First-stage retrieval quality is the clearest moderator. When the
candidate list comes from BM25, PAGC gives large gains and aggregation
often helps. When the candidate list comes from a stronger dense retriever
such as E5, the marginal gain from reranking becomes much smaller.
Aggregation usually matches GCCP-alone and can even hurt on
DBPedia-Entity. This suggests that anchor-based reranking is most useful
when the first-stage retriever leaves enough ranking errors for the
anchor comparison to correct.

The anchor construction shows a similar pattern. Spectral MDS is the most
engineered part of the method, but it is not load-bearing in our
experiments. Simpler anchors built from top-ranked passages or interleaved
top-ranked sentences match or outperform it across our ablations. Thus,
the method works, but under narrower conditions than the original
evaluation suggests. Its main value comes from contrastive scoring, not
from the full set of design choices around aggregation and spectral anchor
construction.

\smallskip

\noindent\textbf{Practical guidance.}
For practitioners, the main lesson is to choose the reranking setup based
on the strength of the first-stage retriever. If the first stage is BM25
or another sparse retriever, 2-component PAGC is worth trying. In this
regime, the candidate list is noisier, the reranker has more room to
improve it, and aggregation gives reliable gains in several settings.

If the first stage is a strong dense retriever such as E5 or BGE, the
picture changes. In our experiments, GCCP-alone is usually competitive
with PAGC. The aggregation step adds extra engineering cost but no
consistent benefit, and it can hurt on entity-heavy datasets such as
DBPedia-Entity. In this regime, practitioners should not assume that the
full PAGC pipeline is better than the simpler contrastive scorer.

The anchor choice can also be simplified. In our experiments, spectral
MDS is not worth the added complexity. A top-3 sentence-interleaved
composite is simpler, faster, and stronger in aggregate on the datasets we
test. Practitioners should also tune the spectral hyperparameters $m$ and
$\theta$ if they use spectral MDS at all. These choices introduce about
$\pm1.5$ points of sensitivity, which is large relative to many reported
state-of-the-art gains. The choice of $z{=}10$ is less fragile because
longer anchors are mostly truncated by the encoder's 512-token limit.

Finally, the method is not tied to encoder-decoder models. The mechanism
also transfers to decoder-only LLMs, including a 4-bit AWQ-quantized 72B
model on a single 48~GB GPU. However, our results suggest that backbone
family matters more than parameter count at the 7--8B scale.

\smallskip

\noindent\textbf{Implications for IR evaluation methodology.}
Our statistical analysis shows why component claims need conservative
testing. The contrastive signal is positive in direction across 19 of 22
settings, but it is Holm-significant as a per-setting effect in only 3 of
22. This does not mean the signal is useless. Rather, it shows the
difference between a broad directional trend and a reliably detectable
effect in an individual dataset.

This distinction matters because IR papers often report many per-cell
significance tests across datasets, models, and ablations. Without
multiple-comparison correction, this practice can make component effects
look stronger than they are. In our results, uncorrected testing suggests
that PAGC improves over GCCP-alone in 8 of 22 settings. After
Holm-Bonferroni correction, only 5 remain significant, including one case
where PAGC is significantly worse. The lesson is not specific to
GCCP/PAGC. Any paper that reports many ablations across multiple datasets
without correction risks overstating which components are truly reliable.

For this reason, we release per-query scores. This allows readers to
repeat the analysis with other correction procedures, including
Benjamini-Hochberg correction or a single joint correction across all
comparison families.

\smallskip

\noindent\textbf{Implications for IR reproducibility.}
The reproduction process also shows how fragile modern neural ranking
pipelines can be. We identify eight operational choices that separate a
paper-only reimplementation from a faithful working pipeline. Several of
these choices fail silently: the code runs, the outputs look plausible,
but the scores are wrong. Target-token case, decoder input formatting,
score normalization, sentence segmentation, and the nDCG implementation
are not minor details in this setting. They can change the conclusion of
an experiment.

This is not a criticism of one paper alone. It reflects a broader problem
in IR reproducibility. Modern reranking pipelines contain many small
implementation decisions that are often not fully specified in the paper
text. Released code helps, but code alone is not always enough. It still
needs to be audited, interpreted, and connected back to the claims in the
paper.

The broader lesson is that documentation norms need to match the
complexity of current neural ranking systems. Papers should specify
prompt formats, target tokens, score normalization, input construction,
candidate-field handling, sentence segmentation, and evaluation
implementation. These choices are easy to omit, but they can be
load-bearing. By releasing our catalogue of operational choices and
per-query scores, we aim to make the reproduced result easier to verify,
stress-test, and build on.
\section{Conclusion}
\label{sec:conclusion}

We studied when anchor-based pointwise LLM reranking helps, using
GCCP/PAGC~\cite{long2025precise} as a representative method. Our
reproduction closely matches the reported results, but also reveals that
the pipeline depends on several undocumented operational choices. These
choices matter: some fail silently and can change the outcome of the
experiment.

Our controlled analyses show that the core contrastive idea is useful,
but its benefits are more conditional than the original evaluation
suggests. Adding anchor-based contrastive scoring to standard pointwise
grading gives robust gains after statistical correction. However, the
aggregation step helps mainly with BM25 first-stage retrieval and adds no
reliable benefit with stronger dense retrievers such as E5 and BGE. The
spectral MDS anchor is also not necessary. Simpler anchors built from
top-ranked passages or sentences match or outperform it in our
ablations. These findings hold across backbone families.

Overall, anchor-based pointwise reranking is effective, but not every
part of the original pipeline is load-bearing. Its gains depend on
first-stage retriever quality, aggregation strategy, anchor construction,
and evaluation protocol. More broadly, our study shows the value of
reproduction-centered stress testing: reliable progress in IR requires
knowing not only whether a method works, but when it works, why it works,
and which components are actually responsible.

\begin{acks}
We thank Long et. al \cite{long2025precise}, the original authors, for releasing their codebase, which was essential for auditing the
implementation choices catalogued in Section~\ref{sec:sensitivity}.
\end{acks}

\section*{GenAI Usage Disclosure}

Generative AI tools were used in the preparation of this work in two ways. First, for language editing of the manuscript, that is, improving the clarity, grammar, and phrasing of text written by the authors.
Second, as a coding assistant when writing and debugging the experimental code. All GenAI-assisted code was reviewed and tested by the authors. No generative AI tools were used in the research design, in the formulation of the research questions, in the generation or processing of data, in the execution of experiments, or in the analysis and interpretation of results. The authors take full responsibility for the entire content of this paper.

\bibliographystyle{ACM-Reference-Format}
\balance
\bibliography{references}

\end{document}